# Transforming the Canada France Hawaii Telescope (CFHT) into the Maunakea Spectroscopic Explorer (MSE): A Conceptual Observatory Building and Facilities Design

Steven E. Bauman*[a], Greg Barrick[a], Tom Benedict[a], Armando Bilbao[c], Alexis Hill[b], Nicolas Flagey[b], Casey Elizares[a], Mike Gedig[d], Greg Green[a], Eric Grigel[e], David Lo[d], Ivan Look[a], Thomas Lorentz[c], Nathan Loewen[f], Eric Manuel[e], Alan McConnachie[b], Ronny Muller[d], Gaizka Murga[c], Rick Murowinski[b], Federico Ruan[e], Derrick Salmon[b], Kei Szeto[b], Jose Teran[e], Rafael Urrutia[c]

[a] CFHT Corporation, 65-1238 Mamalahoa Hwy, Kamuela, Hawaii 96743, USA
[b] MSE Project Office, 65-1238 Mamalahoa Hwy, Kamuela, Hawaii 96743, USA
[c] IDOM Consulting, Engineering, Architecture, Zarandoa Etorb., 23, 48015 Bilbo, Bizkaia, Spain
[d] Dynamic Structures Ltd, 1515 Kingsway Avenue, Port Coquitlam, BC, Canada V3C 1S2
[e] M3 Engineering & Technology Corporation, 2501 W Sunset Road, Tucson, AZ 85704, USA
[f] Sightline Engineering, Vancouver, BC, Canada

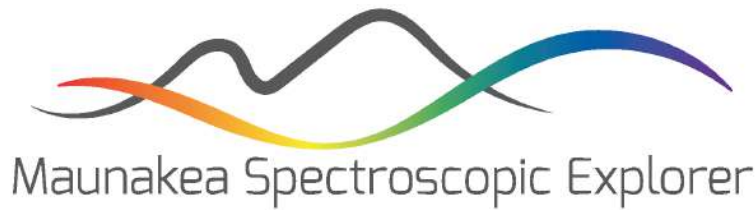

## ABSTRACT

The Canada France Hawaii Telescope Corporation (CFHT) plans to repurpose its observatory on the summit of Maunakea and operate a new wide field spectroscopic survey telescope, the Maunakea Spectroscopic Explorer (MSE). MSE will upgrade the observatory with a larger 11.25m aperture telescope and equip it with dedicated instrumentation to capitalize on the site, which has some of the best seeing in the northern hemisphere, and offer its user's community the ability to do transformative science. The knowledge and experience of the current CFHT staff will contribute greatly to the engineering of this new facility.

MSE will reuse the same building and telescope pier as CFHT. However, it will be necessary to upgrade the support pier to accommodate a bigger telescope and replace the current dome since a wider slit opening of 12.5 meters in diameter is needed. Once the project is completed the new facility will be almost indistinguishable on the outside from the current CFHT observatory. MSE will build upon CFHT's pioneering work in remote operations, with no staff at the observatory during the night, and use modern technologies to reduce daytime maintenance work.

This paper describes the design approach for redeveloping the CFHT facility for MSE including the infrastructure and equipment considerations required to support and facilitate nighttime observations. The building will be designed so existing equipment and infrastructure can be reused wherever possible while meeting new requirement demands. Past experience and lessons learned will be used to create a modern, optimized, and logical layout of the facility. The purpose of this paper is to provide information to readers involved in the MSE project or organizations involved with the redevelopment of an existing observatory facility for a new mission.

*bauman@cfht.hawaii.edu; Telephone: 808-885-3172

# 1. INTRODUCTION

The Maunakea Spectroscopic Explorer (MSE) is a project to upgrade the existing Canada-France-Hawaii Telescope (CFHT) observatory to a dedicated spectroscopic facility by replacing the existing 3.6m telescope with an 11.25m segmented-mirror telescope. It will be equipped with fiber-fed multi-object spectrographs with reconfigurable fiber inputs at prime focus. The scientific need for a dedicated 10m class wide-field multi-object spectroscopic telescope facility has been desired by the international astronomical community for more than a decade.

In response, the MSE project office plans to repurpose the existing CFHT observatory, retrofitting the inner pier, foundation and outer building to become a new facility while retaining the existing architectural footprint. A new enclosure will be supported by the current outer ring foundation and building framework, and a new telescope structure will be supported by the current concrete inner pier, both of the telescope and enclosure support configurations the same as the current CFH telescope and dome. The observatory infrastructure, facilities, and architectural layout which are part of the MSE Observatory building and facilities (OBF) work package will create a modernized and improved facility design to support the new science mission, meet the latest building codes, and capture critical plant equipment.

The project will modify the upper portion of the existing inner pier which is currently supporting a 3.6M equatorial telescope to accommodate a new larger aperture 11.25m ALT/AZ telescope. The modifications will include removing 2-3ft of the existing top section of the pier to establish a new interface for the telescope azimuth ring track, fortifying the inner pier with carbon fiber to support a pintle bearing support structure and reducing the size of all door openings on the outer diameter of the pier with carbon fiber wrap to increase load capabilities.

MSE will also upgrade the outer building structural steel framework to support a new Calotte style enclosure with a shape and size nearly identical (within 10%) to the current CFHT dome. Structural support upgrades to the outer building are needed to meet seismic code requirements, and, to satisfy that prerequisite, an innovative structural solution called the buckling restraint brace (BRB) will be used to replace the existing lateral steel members, adding a compression capacity which is equal to the tension capacity. The lateral buckling restraint bracing will replace the current $1^{st}$, $2^{nd}$, and $3^{rd}$ floor perimeter bracing of the outer building steel structure. Additional radial bracing will be added to the existing steel framework to accommodate and support the new azimuth ring girder for the Calotte enclosure. And a new thermal management strategy will be developed to control heat loads and efficiently maintain the observing environment temperature inside the enclosure with an overarching goal to minimize thermal induced seeing.

# 2. SITE SOIL CAPACITY

## 2.1 Background

The feasibility of the MSE project critically depends on knowing the bearing capacity of the soils under the foundations of both the inner concrete pier and the exterior outer ring building foundation which are actually two separate structures. To know what type of upgrade is possible, size and mass limitations for the new telescope and enclosure must be known. Therefore, it is essential for the fundamental question to be answered – 'What load can the soils under our facility support?'.

CFHT expected that guidance in this regard would be provided by the initial soils studies carried out in 1973 prior to the initial CFHT construction. Accordingly, we hired an expert local soils engineer, Larry Rapp of AECOM (formerly URS) who is familiar with Maunakea, to review the study and provide answers. However, in studying the report, AECOM found that the 1973 report is lacking certain critical details consistent with current design practices. In order to provide CFHT with the allowable soil capacity, AECOM recommended that two soil core borings be drilled to a depth of at least 30 feet below the exterior grade, or 20 feet below the base of the foundations, whichever is deeper, so that AECOM can confirm the current condition of the underlying cinder soil at the CFHT observatory site and develop the needed design recommendations.

| Sample No. | Depth (ft) | Moisture Contents (%) | Dry Density (pcf) | Direct Shear C (psf) | Direct Shear Ø (Deg) | Liquid Limit | Plasticity Index | Gradation Gravel (%) | Gradation Sand (%) | Gradation Silt/Clay (%) | pH | Min. Elect. Resistivity (ohm-cm) | USC |
|---|---|---|---|---|---|---|---|---|---|---|---|---|---|
| 1-1 | 1.0 | 9 | 75 | | | | | | | | 6.4 | 18,000 | |
| 1-2 | 3.0 | 12 | 86 | | | | | | | | | | |
| 1-3 | 5.5 | 15 | | | | | | 34 | 57 | 9 | | | SW-SM |
| 1-4 | 8.5 | 15 | | | | | | | | | | | |
| 1-5 | 13.5 | 20 | | | | | | 24 | 71 | 5 | | | SW |
| 1-7 | 23.5 | 33 | | | | | | | | | | | |
| 1-8 | 28.5 | 24 | | | | | | | | | | | |
| 2-1 | 1.0 | 8 | 75 | | | | | 33 | 60 | 7 | | | SW-SM |
| 2-2 | 3.0 | 10 | | | | | | | | | | | |
| 2-3 | 5.5 | | | | | | | | | | 6.4 | 17,000 | |
| 2-4 | 8.5 | 12 | 80 | | | | | | | | | | |
| 2-5 | 13.5 | 22 | | | | | | | | | | | |
| 2-6 | 18.5 | 21 | | | | | | 35 | 58 | 7 | | | SW-SM |
| 2-8 | 28.5 | 18 | | | | | | | | | | | |

Table 1: Summary of laboratory test results for boring holes

## 2.2 Farwell subsurface investigation summary

The summit site soils report by AECOM summarizes the findings, conclusions and presents the geotechnical recommendations for the design and construction of the MSE facility. The scope included: 1) drilling and test borings near the building location, 2) laboratory testing to determine the general soil properties, 3) evaluating the soil characteristics, 4) geotechnical engineering recommendations, and 5) final findings report. The investigation was performed to estimate the allowable bearing pressure of the existing foundations.

The borings indicate that the site is generally underlain by about 9 to 11 feet of fill over volcanic cinders which extend to the bottom of both test borings at depths of 30 feet below the existing ground surface. Groundwater was not encountered in either of the borings drilled during this investigation and should not be a factor in the proposed construction.

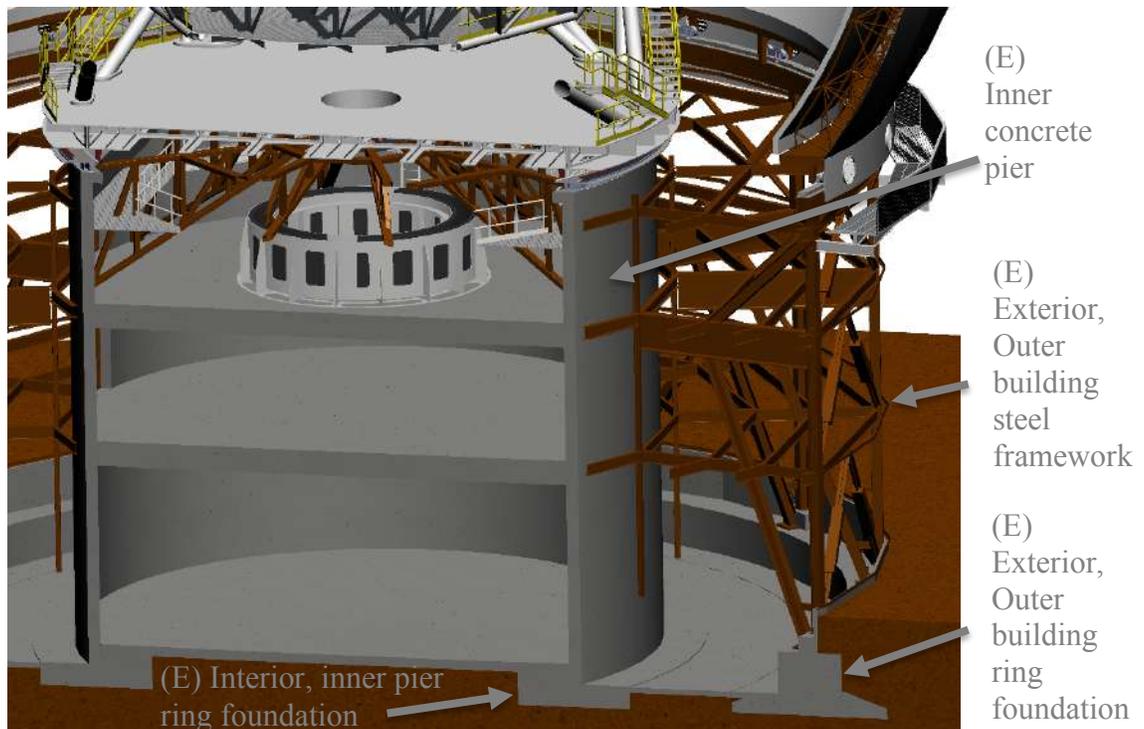

Figure 1: Cross section of the MSE observatory and existing foundations and building steel structure

Based on the relative density of the cinders, and the widths and embedment depths of the existing foundations shown on the plans provided, they concluded that the existing 14-foot wide perimeter ring foundation (outer building) should be able to support an allowable bearing pressure of up to 6,600 psf. The 7.4-foot wide interior ring foundation (Inner pier) which supports the existing telescope should be able to support an allowable bearing pressure of 4,500 psf. <u>The estimated allowable bearing pressure of 6,600 psi (enclosure) and 4,500 psi (telescope) above include a factor of safety of 3 above the foundation bearing failure point based on strength of the concrete and rebar.</u>

MSE will follow the recommendations and <u>limit the foundation contact pressure to no more than 4,500 psf</u> for both the perimeter ring foundation (outer building) which supports the dome and for the interior ring foundation (inner pier) which supports the telescope pier. The allowable bearing pressure may be increased by 1/3 for short-term wind and seismic loads. A friction factor of 0.50 may be used when evaluating the lateral resistance of the foundations. Should the foundation bearing pressures for the upgraded telescope exceed 4,500 psf then additional remedial measures will be taken by MSE to not exceed the geotechnical report recommendations.

## 3. TELSCOPE SUPPORT PIER UPGRADES

### 3.1 Background

This section describes the structural modifications that need to take place in order to meet structural building code requirements for seismic events and reinforcements required to support a new telescope mass with a circular azimuth ring footprint.

### 3.2 Inner pier foundation stability

Fortunately, for the estimated MSE telescope mass, the inner pier foundation can support the load and does not exceed the defined allowable soil bearing capacities. The foundation was also found to be stable for resisting sliding forces and resisting overturning moment from the telescope weight. The wall footing capacities of the pier do not exceed the defined allowable bending moment or wall shear capacities. Essentially the inner concrete pier has no fundamental or structural stability problems and will not require any upgrades or modifications to support a new telescope.

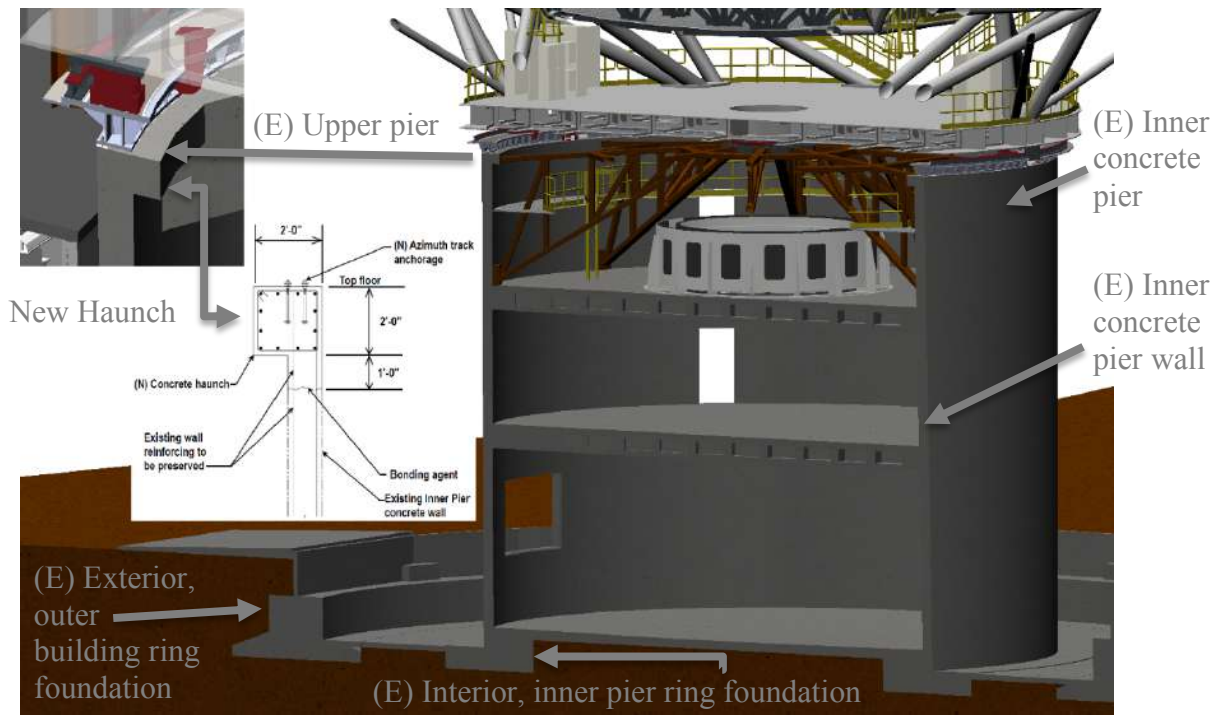

Figure 2: Image of the foundation and inner concrete pier with existing (E) and new (N) upgrades and view of the inner pier wall at the top showing the installation of a new concrete haunch

### 3.3 Structural reinforcements

The existing door opening in the inner concrete pier used to access the CFHT coating facility exceeds the shear stress code limit for un-cracked strength characteristics. Therefore, the size of the opening will be reduced and carbon fiber reinforcement will be added to increase the shear stress capacity. Consequently, carbon fiber reinforcement will also be used at <u>all</u> door openings in the existing pier to increase their shear stress capacity.

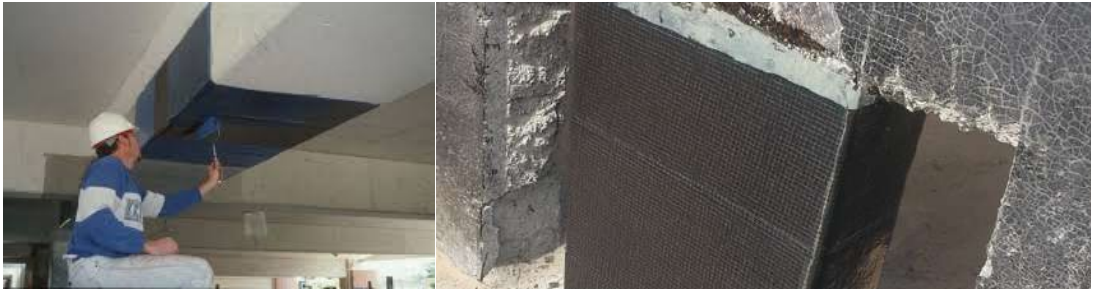

Figure 3: Examples of carbon fiber reinforcement, installation (left) and application (right).

### 3.4 Structural modifications

To accommodate a new ALT/AZ telescope azimuth track with a different circular footprint than the existing A-frame equatorial mount of the CFHT telescope, the top 2-3 feet of the inner concrete pier wall will be removed to facilitate the installation of a new 24inch x 24inch concrete haunch (an extension or knee like protrusion of the foundation wall) with cast-in-place reinforcement to support and anchor the new azimuth track. The existing wall will be removed using temporary reinforcements to preserve the existing concrete structure below and then exposed rebar and bonding agents will be used to add a new 2' wide x 2' tall concrete haunch with concrete pilasters (anchor bolts cast and embedded in the concrete before it dries).

## 4. BUILDING MODIFICATIONS

### 4.1 Background

This section describes the structural modifications that need to take place in order to meet building code requirements for wind and seismic events and discusses the reinforcements required to support a new Calotte enclosure.

### 4.2 Structural Modifications and Retrofits (New)

The figure below shows the structural members of the outer building steel framework. This color coded image will serve as a roadmap in the next sections to highlight and identify the members that require modification, retrofit or do not require any change.

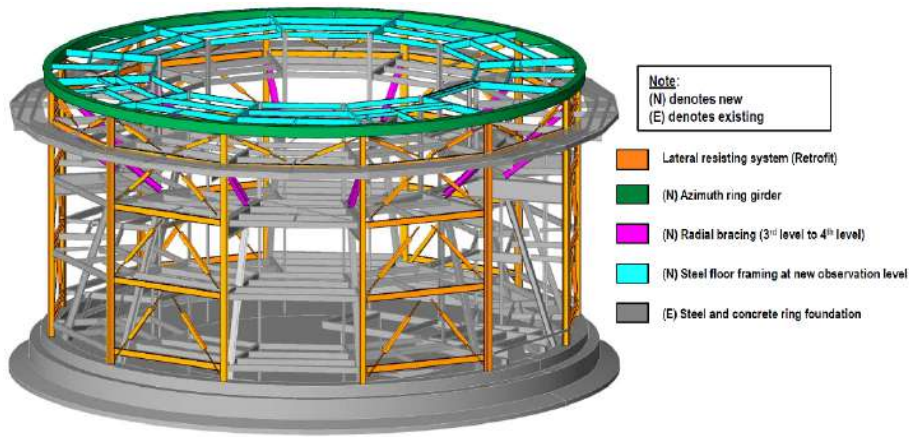

Figure 4: Isometric view of the outer building steel structure with a color coded legend of elements

The new enclosure azimuth ring girder which will support the Calotte enclosure does not exceed the allowable unity ratio (measure of how much of the member's capacity is being used) required by building code for dead load and wind load combinations. Therefore, the design baseline for the MSE structure will be similar to the current CFHT ring girder and will use a steel box type girder with 50ksi (yield) ½ inch steel plate. The size of the girder will be roughly 24inches deep x 14inches wide. The old CFHT dome azimuth ring girder will be replaced by a new MSE enclosure azimuth ring girder but physically at a different location.

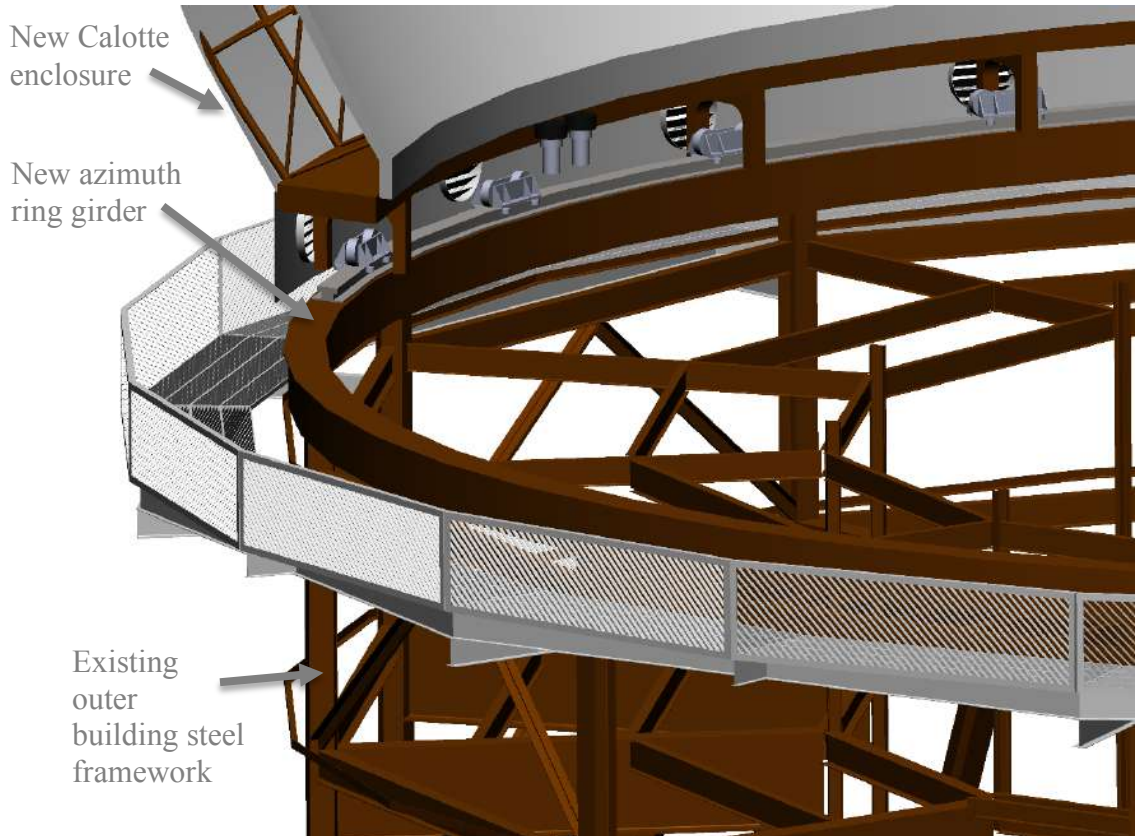

Figure 5: Isometric selective cutaway view showing the new azimuth ring girder for the Calotte enclosure

The new MSE azimuth ring girder and new observation level floor framing require additional diagonal radial bracing to provide additional stiffness and radial restraint resistance to lateral deformations. Luckily, the allowable unity ratio required by code for the load combinations is not exceeded. New radial brace members will be installed and will be connected from the existing 3rd floor girder and will connect to the existing 4th floor framework under the new enclosure track ring girder.

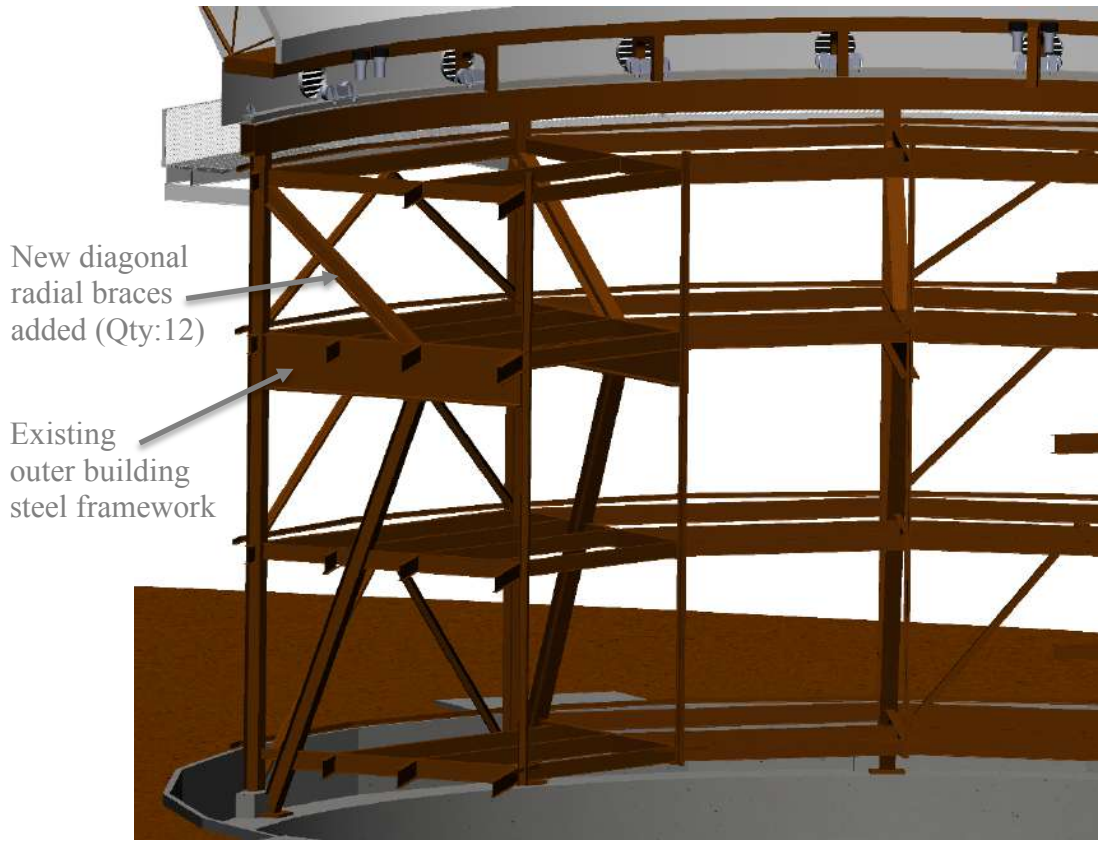

Figure 6: Cross section view of the outer building steel structure showing the location of the new diagonal radial braces

The new MSE observation level floor does not exceed load capacities or deflections allowable required by code. Therefore, the floor will be supported on steel floor beams with a layout and orientation similar to all existing floor framing in the current building. The new floor will consist of 3/8-inch-thick steel plate with L4x3x5/16 steel stiffeners. The uniform live load rating for the floor will be 150 psf.

### 4.3 Structural Additions and Reinforcement (existing)

The existing vertical exterior main support columns and horizontal exterior beams of the outer building steel framework exceed the unity ratio allowable by code. However, the columns and beams can be reinforced by adding 3/8inch steel plates welded to both sides of the W prolife of the flange to increase the capacity and satisfy the building code. This upgrade will be required on all columns and beams designed as part of the new integrated lateral resisting system.

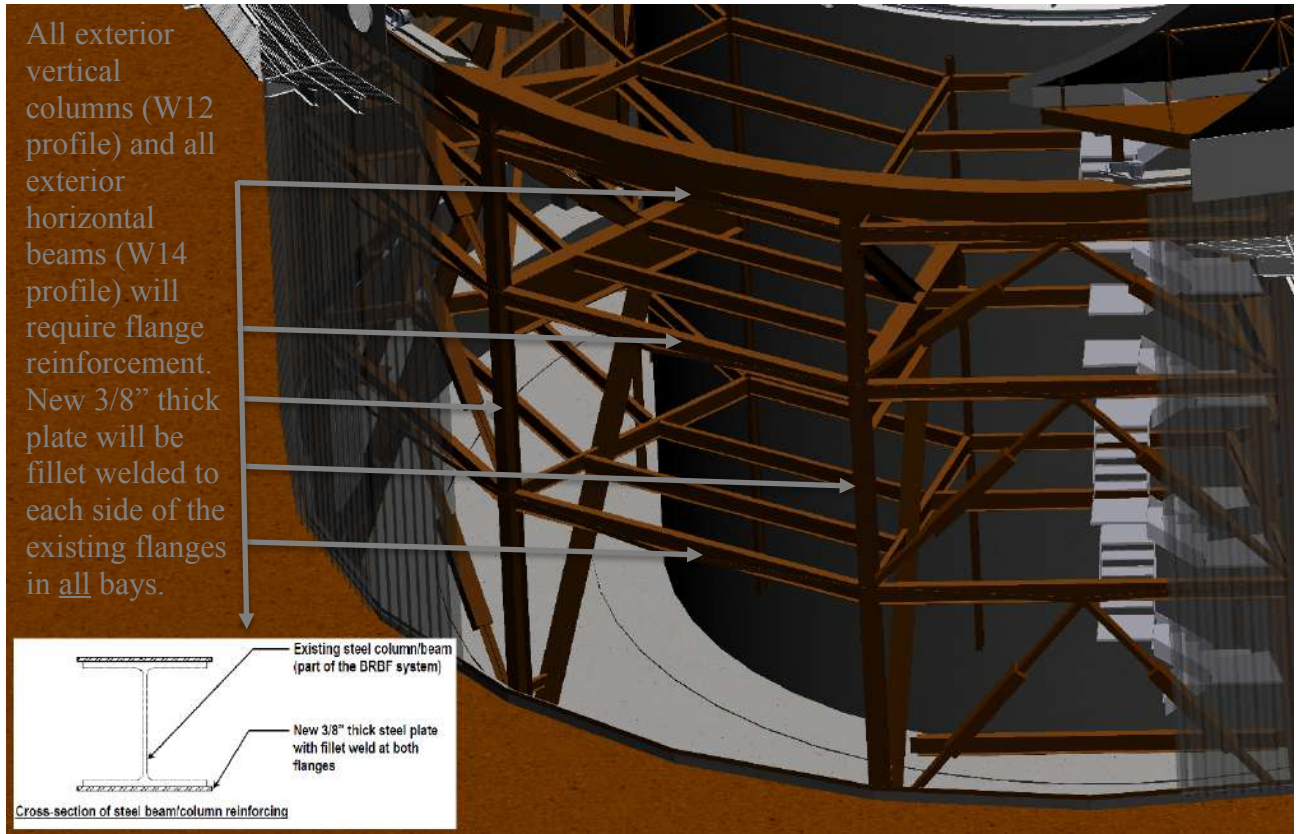

All exterior vertical columns (W12 profile) and all exterior horizontal beams (W14 profile) will require flange reinforcement. New 3/8" thick plate will be fillet welded to each side of the existing flanges in <u>all</u> bays.

Figure 7: Upgrades required to the exterior main support columns and exterior beams.

The existing steel bracing (single angle and double angle) between the exterior main columns in every other angular bay will need to be replaced with an inverted chevron buckling-restrained brace (BRB) to increase the allowable unity ratio required by code. Every bay will contain the chevron bracing between the 1st floor to the new observation level floor and become the main element in the updated new lateral resisting system. The BRB's are a unique structural member since they have equal capacity in compression as well as in tension and dissipate seismic energy via friction.

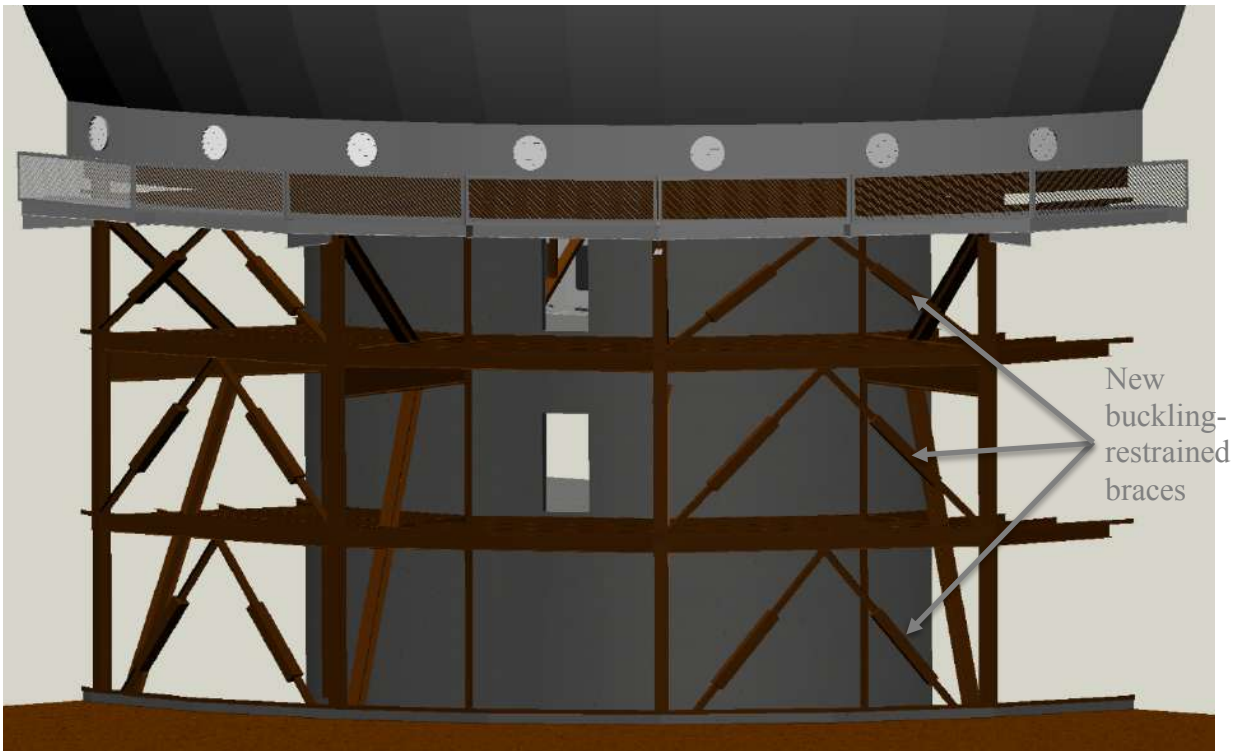

Figure 8: Quarter section view of upgrades required to the angular bays of the outer building steel framework

The existing interior support columns do not exceed the allowable unity ratio required by code for the load combinations and will require no change. So, the existing columns will be trimmed to fit the level below the new azimuth ring girder at the new observation level floor.

The existing tension hanger support columns also do not exceed the allowable unity ratio required by code for the load combinations therefore no changes will be required.

The existing radial support columns do not exceed the allowable unity ratio required by code for the load combinations. Thus, no changes will be made; these members provide a vertical support of the 3rd floor steel girder. The brace is connected to the existing 3rd floor girder and is supported at the base of the concrete foundation

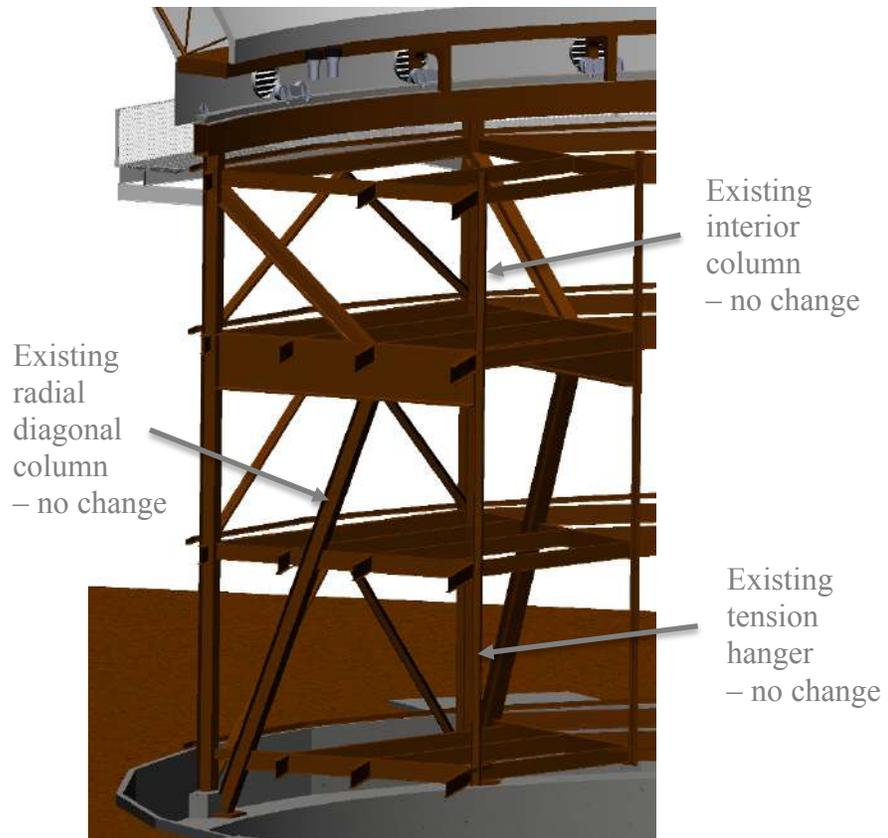

Figure 9: Quarter section view of interior members of the outer building steel framework that require no modification

The existing concrete outer ring foundation is stable for resisting overturning moments; however, during extreme winds, the net uplift at the existing exterior column to concrete foundation interface exceeds the capacity of the existing connection. Therefore, new anchors and a concrete stem wall will be added in place to reinforce the column anchorage at the base of the concrete foundation connection. This update will be required at twelve places where the main columns anchor to the concrete foundation. Holes will be drilled into the concrete and threaded reinforcing bars will be inserted into the existing concrete using an epoxy adhesive system. The intent is to encase the existing connection with concrete and add shear studs to the column to resist the net wind uplift.

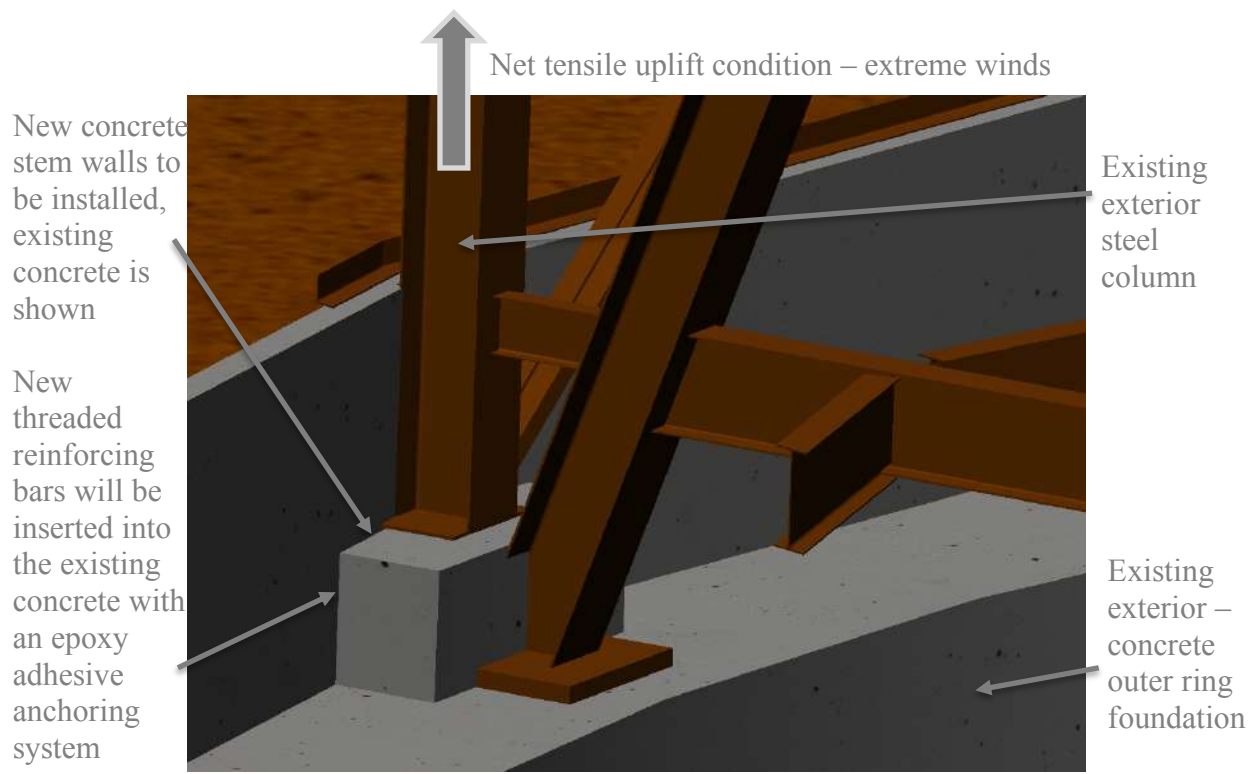

Figure 10: View of the outer concrete ring foundation and exterior column with the proposed structural modifications

## 5. MSE OBSERVATORY OVERVIEW

### 5.1 Overall Architecture layout

The OBF building layout is divided into the outer building levels: basement (bottom level) to the observing floor (top level) and the inner pier levels: coating facility (bottom level) to the top of the pier (telescope azimuth ring support structure).

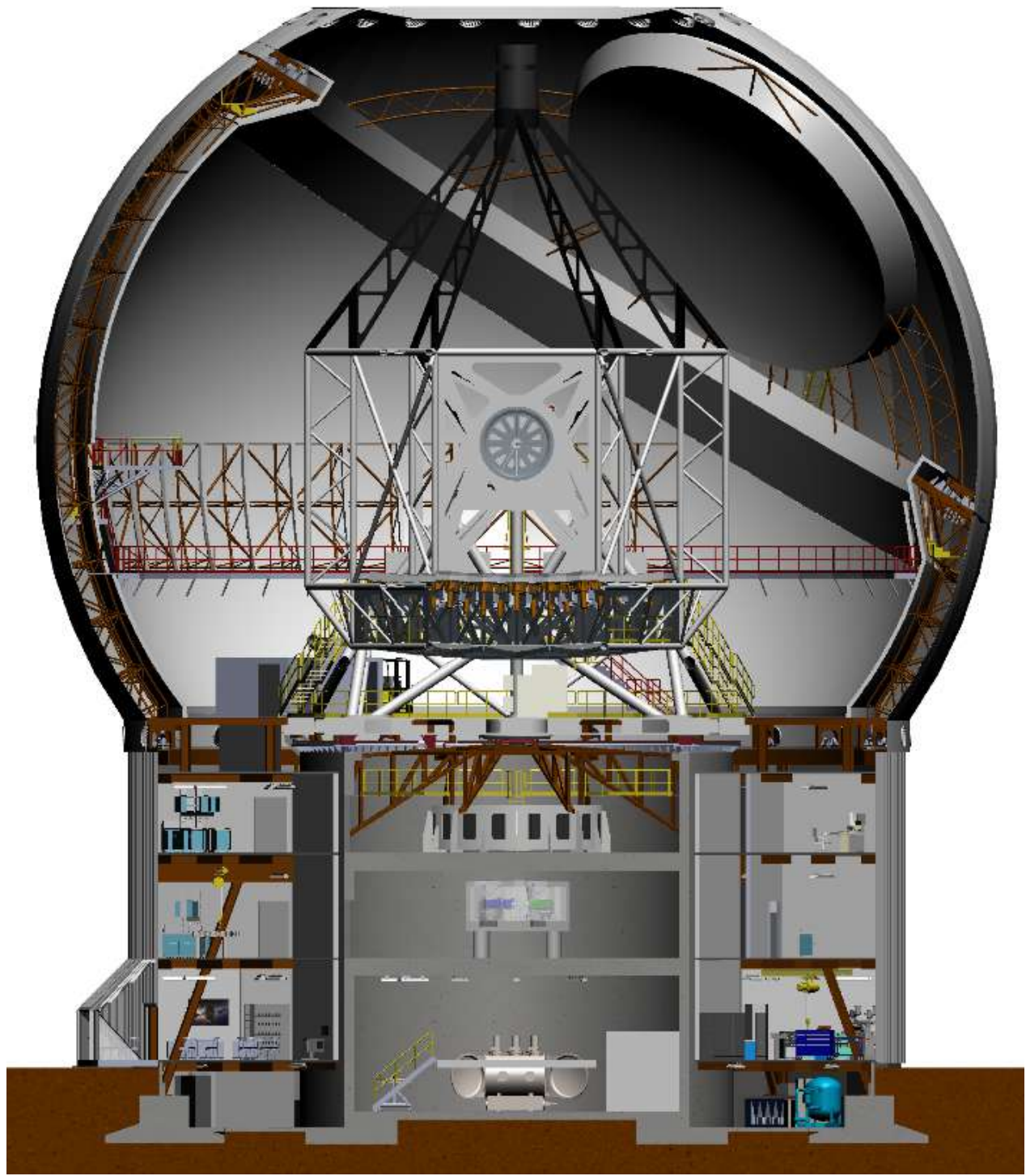

Figure 11: Cross section of the MSE observatory depicting the various levels and systems inside

# 6. OBF CONCEPTUAL DESIGN OVERVIEW

## 6.1 Overview

The observatory building and facilities (OBF) conceptual design includes the engineering studies, requirements, and design modifications of the support structures for the MSE enclosure, MSE telescope, and the infrastructure and equipment required to operate, support and facilitate daytime and nighttime operations.

The previous sections of this paper consisted of the design changes and upgrades needed to alter the existing outer building structure (consisting of five levels of steel framing) and connections to the outer concrete ring foundation to support a different enclosure while meeting building codes, specifically wind loads and seismic loads. It also discussed how the existing inner concrete pier will be changed to support a new telescope with a circular footprint while again meeting seismic and building codes.

The next section will touch on the design changes and modifications of the existing observatory infrastructure, how the layout and organization will transform, and provide details about the equipment (new and reused), where it is located, and what service it provides.

## 6.2 Basement level - Conceptual design (Building)

The bottom level of the basement will be used primarily to house mechanical and electrical plant equipment and utilities, the infrastructure, required to supply and distribute power and cooling to the telescope, enclosure, instrumentation, and building equipment. For optimal thermal management, the largest heat generators will be deliberately placed in the basement, furthest away from the observing level where waste heat can be isolated and managed. More on the thermal management system will be covered in section 7. Only a few pieces of equipment will be housed on the first floor due to other constraints which will be discussed later. Much of the equipment that generates large vibrations is also deliberately located in the basement to distance equipment as far as possible from the telescope and isolate it from the inner concrete support pier.

Some of the plant equipment and utilities include: the domestic water system, main supply building power, main electrical power distribution system, uninterrupted power supplies (UPS), compressed dry air system, building and instrument chiller systems, cryogenic systems, helium compressors, building exhaust system (air handlers and ducting), elevator equipment rooms, a lift platform from the basement to the first level, motion controllers and drive electronics. For environmental protection, fluid containment systems using berms will be installed for spill prevention under equipment.

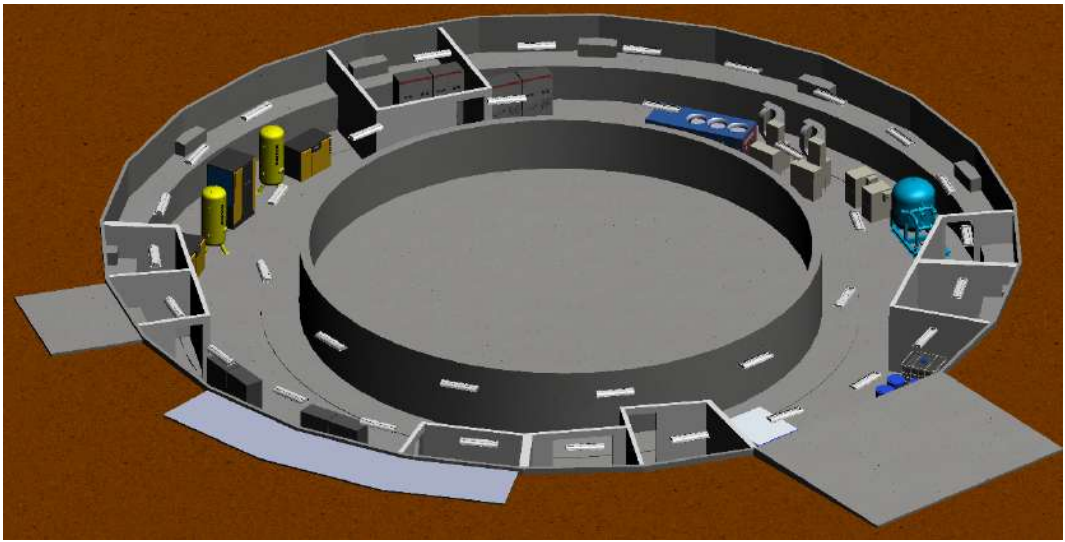

Figure 12: Infrastructure and layout of the basement (bottom level - outer building) with the inner pier contents removed.

### 6.3 Coating facility - Conceptual design (Inner pier)

The bottom level of the inner pier, comprised of two levels with the ceiling height of the basement and first level combined, will remain a mirror coating facility. However, the majority of the CFHT infrastructure and equipment will be removed to allow upgrades. The new MSE coating facility will be a space dedicated to mirror segment coating, servicing, and safe storage. The coating facility will contain varying clean-room levels for various activities, such as the removal of old coatings, cleaning and final prep for new coatings, and vacuum coating chambers, along with the appropriate sized roughing pumps, turbo molecular pumps, and power equipment. The facility will also contain designated storage areas for segments in process and off-telescope spare mirror segments. The coating facility will provide handling and lifting equipment and segment handling carts for safe transportation of segments to and from the observing level. A lift platform will be added to the coating facility to safely lift/lower segments on handling carts from the first level of the outer building to the bottom floor of the coating facility. The existing polar crane could be reused and upgraded with new motors, reduction units, and motor controllers as needed. Any other equipment that can be repurposed will be used accordingly.

### 6.4 Ground level - Conceptual design (Building)

The main entrance and first level of the observatory will be designed and arranged to support staff and visitors' needs. Just after the front entrance is a small mail room, an area for hand held radios, and an observatory status screen providing feedback status on all observatory systems and equipment. To the right inside the main entrance are restrooms and a common area with a kitchenette and lounge. To the left of the main entrance is the telescope hydraulic power unit (HPU), backup power generator, above ground storage fuel tank, and fabrication shop.

The common area has the following, a first aid treatment room outfitted with a cot, AED, medical oxygen and first aid medical equipment and supplies. The lounge is equipped with tables and chairs for eating, furnishings for sitting and relaxing, cabinets and drawers for food storage and cookware, and a kitchenette for meal and drink preparation. There is also a wall area for staff and guests to interface with standing computer consoles and a covered patio area adjacent to the kitchenette for sitting outside. The relocation of the common area from the existing 4th level to the ground level now accommodates staff and visitors' needs immediately upon entry to the observatory. It also locates personnel and their warm bodies (heat sources) away from the observing level.

The fabrication shops are located on the ground level and includes a machine shop, welding shop, material storage area for metal stock, hand tool storage, and hardware and plumbing stock storage area. The main shipping and receiving storage area, with shelving and space to contain large deliveries, is next to the fabrication shop to accommodate ease of restocking inventory of parts and supplies. Keeping the fabrication shop next to the receiving area also supports repair and maintenance of mechanical equipment. The main shipping and receiving area has a large multi-leaf access door to the outside of the building measuring 21ft wide by 24ft high which will be reused for MSE. A new lift platform to reach the 1$^{st}$ floor from the basement below will be added, and a 3-ton jib crane for loading and unloading will be updated and reused. Various handling and lifting equipment will be stored in the receiving area. Much of this equipment already exists and is being updated and reconfigured to improve workflow and logistics.

The HPU (supplied by the telescope structures group), comprised of a fluid reservoir, motors; pumps, filtration, cooler, plumbing, lines, and fittings for the hydrostatic bearing support system will be located on the ground level. The area around the HPU will have berms installed under the equipment as a containment mechanism in case of fluid leaks. The ground level remains the location for the HPU since it provides additional levels of protection before fluids can reach the basement level and minimizes any fluid from encountering the ground. The fluid would have to overflow two concrete berms and each levels floor before reaching the soil. The backup power diesel generator also stays on the first floor to accommodate a new above ground storage tank for diesel fuel which eliminates the old underground storage tank and minimizes environmental and health department regulations. Berms will continue to be used under the diesel generator and tank, and additionally a storage tank with secondary lining will be used to avert fuel leaks from encountering the soil.

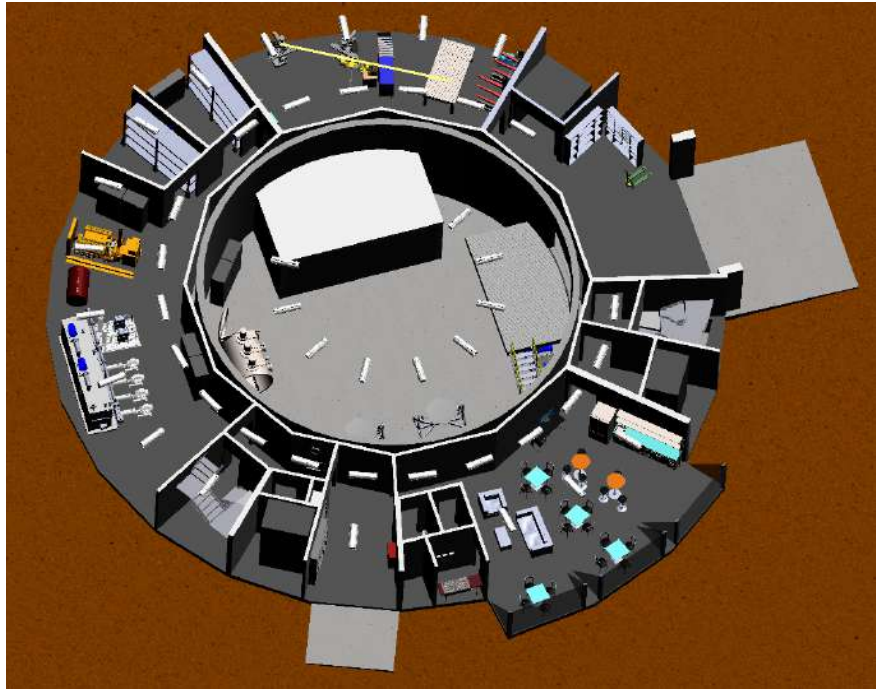

Figure 13: Infrastructure and layout of the ground level of the building and the bottom and ground levels of the inner pier

## 6.5 Second level – Conceptual design (Building)

The second level of the building will provide staff with laboratories, offices, and general working and meeting areas.

The office for the site engineer will have a working area to support day to day operations. A technical documentation library, shelves with paper design documents, engineering drawings and manuals will be co-located next to the office for ease of access. The office will have a large format printer for creating paper copies of digital drawings. A printer/scanner/copy machine will also be available as well as a cabinet with general office supply materials.

A locker room area will provide staff with individual closets for personal items, cold weather gear, personal protective equipment (PPE), boots, and a place to store general belongings.

A conference room will be located on the second level and will accommodate approximately 12 people for general meetings and discussions. It will include modular tables and chairs (stackable and configurable), a digital projector and /or large LED panel television monitor, a simple computer interface, and basic audio and video communications equipment.

A small fire proof flammable cabinet and chemical storage room is envisioned in the observatory; however, it would be better to move these items to the headquarters (HQ) facility and minimize combustibles on site. Similarly, the bulk of the spare mechanical, electrical and hardware parts and spares will be stored offsite at HQ to further reduce flammable materials on site.

The laboratories on the second level will consist of an electronics laboratory, a fiber optic lab, a vacuum/cryogenic laboratory, and a spectrograph laboratory with clean room.

The electronics lab (E-lab) will consist of two electro static discharge (ESD) benches with oscilloscopes and other various electronic test equipment. Each table will have power outlets as an integrated part of the test bench. A tool chest with general hand tools and shelves to store testing equipment will be installed inside the E-lab. Both the fiber optic lab and vacuum cytogenetic lab will be outfitted in a similar fashion and populated with tools and equipment based on the needs of the instruments and scientific equipment inside the observatory.

The clean room; will be large enough to accommodate routine service, minor repairs, testing, and upgrades to the spectrograph instruments and equipment. The clean room with have an overhead ½ ton monorail crane for handling and lifting parts and components. The crane is an existing piece of equipment and will be upgraded as needed to support spectrograph instrumentation and scientific equipment needs once more information is known.

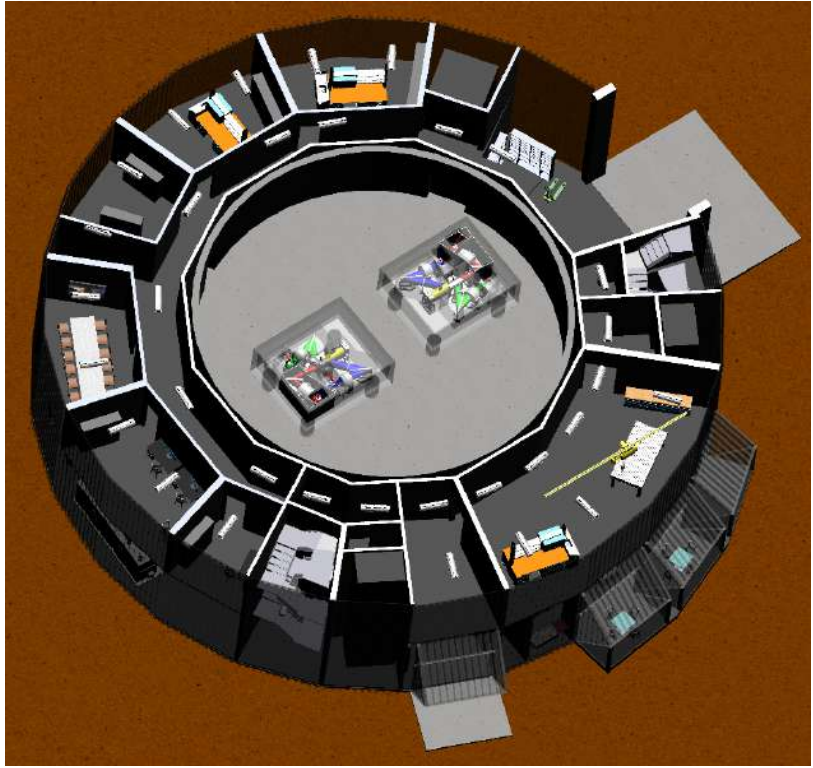

Figure 14: Infrastructure and layout of the second level of the building and second level of the inner pier

### 6.6 Spectrograph room – Conceptual design (Inner pier)

The second level of the inner pier will house multiple fiber feed spectrographs and related support equipment. At this time, during the conceptual design phase, the high resolution spectrograph (HR) will be located in the second level inner pier room. However, if things were to change, the space inside the inner pier would support the low-medium resolution spectrograph (LMR) configuration alternatively.

### 6.7 Third Level – Conceptual design (Building)

Below the observing level and 4th level crawlspace is the third level of the outer building which contains the main computer room, observatory control room, and access to the pintle bearing, cable wrap and telescope azimuth rotational mechanisms inside the inner pier. The observatory control room will be outfitted with workstations to enable the control and operation of the telescope, enclosure, instruments and building systems.

The observatory control room is located adjacent to the main computer (CPU) server room which utilizes an active cooling system to maintain the ideal environment inside the server room for computing systems. Another set of bathrooms are located just a short walk down the hall from the control room. The fire protection system, also located inside the observatory control room, operates and monitors the clean agent fire suppression system located inside the CPU server room. The fire suppression system protects the critical computer systems and communications systems for the observatory in the event of a fire. The room is sealed, to contain the chemical fire retardant and also establishes a constant volume for the cooling system to maintain.

There is undesignated space on the third level that could be used for safety equipment storage, a secondary first aid room with emergency supplies and/or other offices and labs, however these ideas are just placeholders for now. A large amount of space on the third level has purposely been left unassigned to accommodate future developments, unforeseen additions and expansion in the future phases of the MSE project

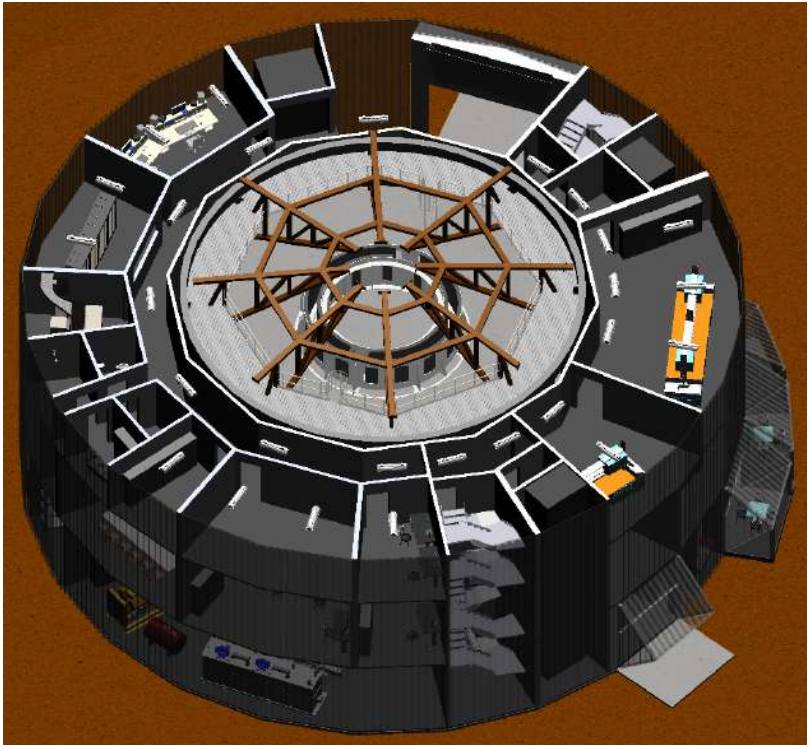

Figure 15: Infrastructure and layout of the third level of the building and the third level of the inner pier

**6.8 Pintle bearing and Cable wrap access – Conceptual design (Inner pier)**

The third level of the inner pier will provide access to the pintle bearing rotating mechanisms, telescope azimuth cables (power, communication, optical fibers, cooling, pneumatic lines, etc.) and the support structure for the bearing assembly and azimuth track lateral and seismic restraints. The inner pier room will also house cabinets with various tools and equipment, tool chests with various hand tools, and a platform system (stationary or movable) for accessing components of the cable wrap and bearing system for repairs, maintenance and upgrades.

**6.9 Fourth level crawlspace – Conceptual design (Building)**

The fourth level of the outer building will have limited overhead height and will not be used during normal work hours; it will normally be void of personnel and have two entrance/exit points. The small volume created from the old CFHT 4th floor level and the underside of the new observing level will create a "squat height" crawl space and serve the main function as an air plenum and secondarily a bypass space for cables, plumbing, and electrical runs. The plan is to use the empty void as a thermal isolation barrier preventing heat from entering the observing space, flushed by passive or active fan systems to remove heat from entering into the observing level. More information related to this can be found in section 7.

## 6.10 Observing Level – Conceptual design (Building)

The observing level of the observatory provides access to the telescope, enclosure, and instrument systems along with handling equipment and tools. The observing level can be entered by using one of the two staircases or by elevators. There are three (3) elevators in the building, and they all terminate with the observing level as the highest point of travel. The staff elevator is intended for daily normal work traffic of staff and visitors, the segment elevator for transport of segment assemblies to and from the coating facility, and the freight elevator to transport equipment, supplies and materials from the levels below to the observing level. Three elevators exceed the building code requirement number; therefore, the number of elevators could be minimized in future design phases. However, the usage for each elevator currently serves an important purpose at this time and will remain part of the design.

Personnel handling equipment such as man lifts, forklifts and articulating boom lifts will be located on the observing level and provided for staff to access the telescope, instruments, and enclosure systems for maintenance and repairs, and to handle parts and equipment as needed. Tool chests and equipment cabinets with maintenance and repair tools and equipment will be located on the observing level as well.

The enclosure will provide multiple cranes to enable staff to lift and transport heavy equipment and instruments within the observatory and on the telescope. Handling fixtures and rigging equipment will be located in storage cabinets on the observing level.

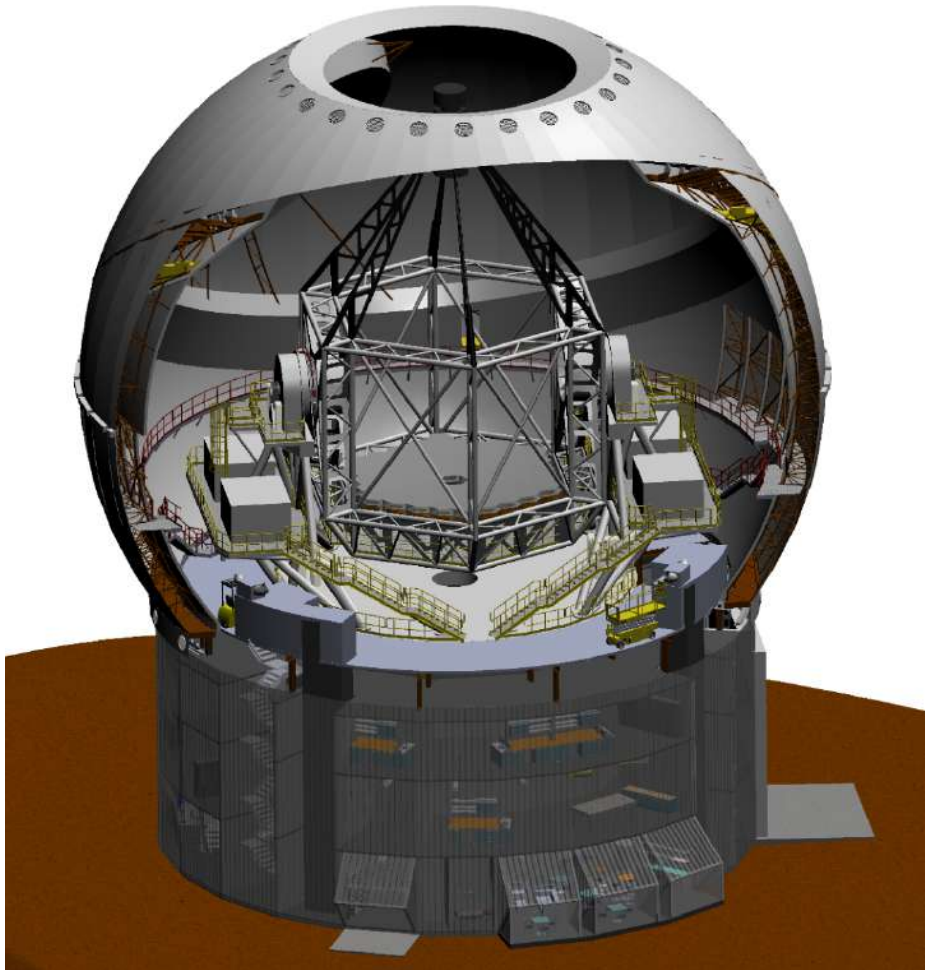

Figure 16: Infrastructure and layout of the observing level of the observatory

# 7. THERMAL MANGEMENT PLAN

## 7.1 Background

The thermal design strategy and plan for MSE outlines steps necessary to minimize the thermal-induced seeing on the telescope and discusses the plan to develop strategies for the MSE building, enclosure, telescope and inner pier in the next phase of the project. The goal of the thermal management system is to ensure that the MSE Image quality (IQ) is as close to the natural site limit as is possible.

## 7.2 Overview

A new thermal management strategy will be developed to control heat loads and efficiently maintain the observing environment temperature with an overarching goal to minimize thermal induced seeing. MSE will utilize a thermal management system based on thermal principals for isolating heat generators, minimizing solar loads, reducing air infiltration, insulating large structures, controlling equipment specific heat loads and flushing daytime and nighttime environments with air plenums to maintain air volume exchanges. For instance, the existing CFHT building exhaust tunnels in the basement leading outside will continue to be used for MSE to remove heat loads deliberately positioned in the basement to minimize thermal loads inside the building. Remaining residual heat that makes its way up from lower levels will be flushed prior to reaching the observing environment using an active and/or passive air plenum system. An active cooling system inside the enclosure will be used to keep large structures and the observing environment near the ideal nighttime observing temperatures.

## 7.3 Thermal strategies

The MSE thermal management system will use conventional engineering approaches to estimate the heat load and mitigate heat dissipation into the interior space of the enclosure above the observatory level where the telescope and instruments are located. Passive thermal management strategies such as insulation and coatings are preferred where possible over active strategies such as forced air induction, which consumes electric power, produces vibrations, and incurs operating costs.

## 7.4 Building strategies

The basement and ground levels of the observatory will house the main facility and utility equipment that are large heat generators and keeps them as far as possible from the observing environment. Some of this equipment includes the main electrical transformers and switchgear, instrument and building chillers, compressors, pumps, cryogenic equipment, electronics, and hydraulic equipment. The ceiling of the first level will be thoroughly insulated to prevent heat transfer to upper floors of the observatory via a thermal insulation barrier. Fresh outside air will be brought in from the outside as make up air to the building exhaust and used to keep the lower two levels cool. These combined lower level heat loads will be removed by a forced air system which rejects heat from the building far from the observatory via two existing underground tunnels. This generates from two tunnels heat plumes as far as possible from the Observatory on the predominantly downwind sides of the summit site. Thus minimizing the negative impacts on telescope performance. Vented louvers on the exterior of the observatory in line with the fourth level crawl space will be used as a passive or active ventilation system to remove any residual heat parasitically leaking from the lower levels and vent it to the outside, creating a final defense and thermal isolation barrier.

## 7.5 Telescope strategies

Heat transfer from the telescope will be minimized by insulating the structure (thermal mass) with heavy duty insulation to reduce thermal convection and by covering the insulation with a reflective surface to minimize thermal radiation. The telescope and instruments will be coated with low emissivity paint to prevent radiative super cooling at night to temperatures below ambient. Heat sources that stay with the telescope will be enclosed in temperature controlled enclosures where chilled water will be used to transfer heat loads via heat exchangers to the basement. The hydrostatic bearing oil will be sub-cooled after the hydraulic power unit to match the nighttime ambient air temperature so when the fluid exits the hydrostatic pads the thermal effect is null.

## 7.6 Enclosure strategies

Like the telescope the enclosure will also be painted with a low emissivity paint which has low solar energy absorption properties to minimize thermal radiation. The interior walls of the enclosure will be insulated and possibly reflective for the same reasons discussed before, but stray light considerations may not allow this. The interstitial space (gap between the exterior and interior walls of the enclosure) will be vented at the top from the bottom to cause an upward flow of air to flush the space and minimize convective heat transfer from the exterior walls. Heat generators that must reside inside the rotating enclosure will be located in the vented interstitial space. Air handlers inside the enclosure volume will circulate chilled air inside the observing environment to achieve a uniform temperature distribution, with the intent of keeping the telescope structure, instruments, building structure and observing level flooring at the anticipated ambient night-time air temperature. In addition, vent doors on the enclosure diameter will provide flushing of the interior observing space to remove hot air and minimize thermal gradients during observing. Static and rotating seals for the enclosure will have sufficient performance to minimize air infiltration and daytime air conditioning loads.

## 7.7 Thermal Management system

The thermal management system design will incorporate all of the methods discussed earlier and the design of these systems will be addressed during the next preliminary design phase. Below is a visual representation of the MSE observatory showing the thermal management subsystems and their locations.

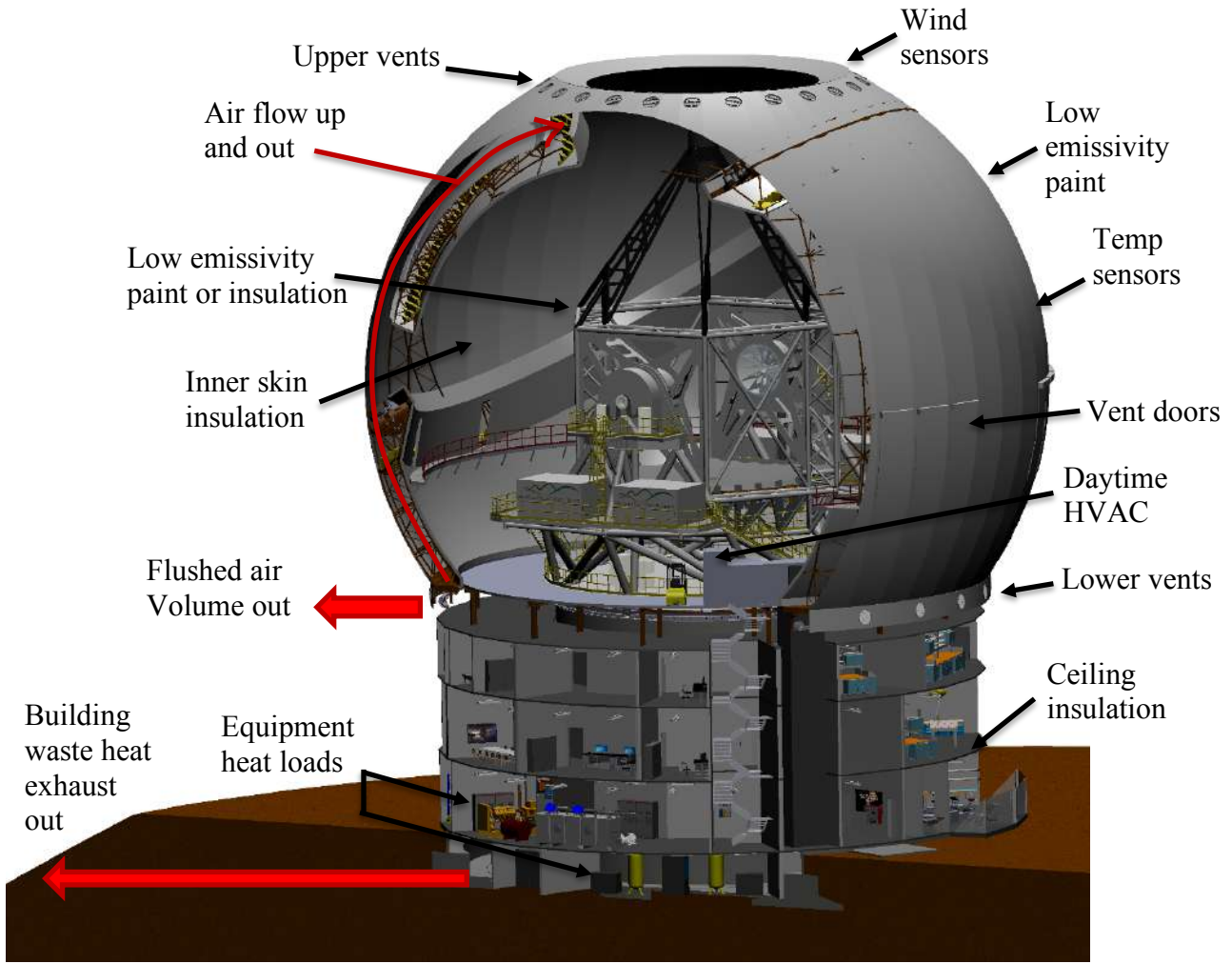

Figure 17: Depiction of the MSE Observatory and the thermal management system components

## 8. SUMMARY

The future plans to transform the Canada France Hawaii Telescope into the Maunakea Spectroscopic Explorer are becoming a reality and with the completion of the OBF conceptual design in later 2018 no show stoppers have surfaced yet. However, the project has overcome many obstacles that could have ended the efforts to replace the 3.6m Equatorial telescope with an upgraded 11.25m ALT/AZ telescope.

The existing telescope interior (inner pier) ring foundation will be repurposed to support a new telescope and the existing perimeter (outer building) ring foundation will be upgraded to support a new Calotte enclosure. Therefore, both foundations need to have adequate capacity to support the mass and footprint of a new telescope and enclosure, while meeting current International building code (IBC) requirements. A geotechnical study was performed on site to determine the bearing capacity of the soil under the building and pier foundations. The report provides the allowable bearing pressures of the foundations which confirms that the foundations can support the MSE telescope and Calotte enclosure plus establishes the mass limits for the telescope and enclosure according to the foundation contact pressures. The inner concrete pier was found to have no fundamental or structural stability problems and will only require minor upgrades to reinforce all door openings and modifications to the upper top portion of the pier to create a section of thicker foundation wall to support and anchor a new circular azimuth track for the MSE telescope.

The existing structural members of the outer building steel framework will require either a retrofit, upgrade, replacement or modification to accommodate a slightly taller telescope and new Calotte enclosure. These changes range from a new azimuth ring girder at a different elevation with new diagonal radial braces added from the $3^{rd}$ floor to the enclosure ring girder structure to structural reinforcements such as adding 3/8-inch steel plate to the flanges of the vertical columns and horizontal beams to increase their capacity to satisfy building codes. Specific steel bracing between main columns in every other angular bay will be replaced with buckling-restrained bracing, a structural member that can dissipate seismic energy with a capacity equal in tension and compression. Finally, the existing outer ring foundation will need to be reinforced at the base of the vertical columns connecting to the concrete foundation to resist net uplift during extreme winds.

The primary function of the OBF is to protect personnel and equipment inside the observatory from the exterior environment, and the second function, is to establish an infrastructure and facilities to enable science observations. The design changes and modifications to the existing CFHT observatory follow the conceptual design requirements for the MSE project and demonstrate the ability to transform the current facility into a new MSE observatory which satisfies the primary and secondary functions. The outer building and inner pier infrastructure and floor plans have been modified and improved to support a new telescope, enclosure, and instrumentation while optimizing general operations and daytime maintenance and repairs.

A thermal management plan has been developed that incorporates strategies to minimize thermal-induced seeing on the telescope which will be pursued in the next phase of the project with an overarching goal of ensuring that the MSE image quality is as close to the natural site seeing as is possible. The future looks promising for the MSE project and with this work completed the next phase of the project can start with a stable foundation with the capacity to support the detailed science case proposed by the MSE project and contributing scientists.

## ACKNOWLEDGEMENTS

CFHT would like to express its gratitude to the companies involved in the MSE project: Dynamic Structures, M3 Engineering, IDOM, Fewell Geotechnical Engineering and Sightline Engineering. Their contributions and support are invaluable and keep CFHT and MSE, competitive in the astronomy community.

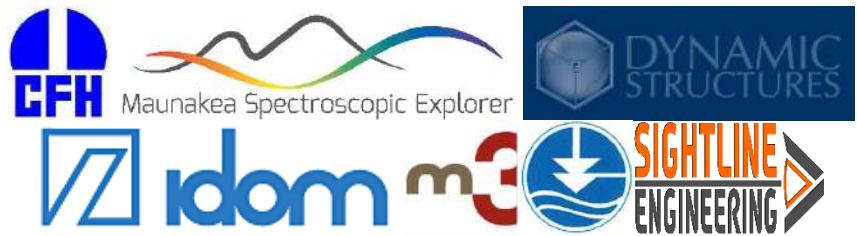